\documentstyle{aipproc}

\input epsf 

\begin{document}

\title{
``Convergent observations'' with the
stereoscopic HEGRA CT system
}

\author{Hubert Lampeitl and Werner~Hofmann \\ for the HEGRA Collaboration}

\address{Max-Planck-Institut f\"ur Kernphysik, P.O. Box 103980,
        D-69029 Heidelberg, Germany}

\maketitle

\begin{abstract} 
Observations of air showers with the stereoscopic HEGRA IACT
system are usually carried out in a mode where all telescopes point in
the same direction. Alternatively, one could take into
account the finite distance to the shower maximum and orient
the telescopes such that their optical axes intersect at
the average height of the shower maximum.
In this paper we show that this ``convergent observation mode'' is 
advantageous for the observation of extended sources and for surveys,
based on a small data set taken with the HEGRA telescopes 
operated in this mode. 

\end{abstract}


The HEGRA collaboration is operating a system of currently five
imaging atmospheric Cherenkov telescopes (IACTs) for the 
stereoscopic observation of VHE cosmic $\gamma$-rays
\cite{performance}. The telescope system is located
on the Canary Island La Palma, at the Observatorio del
Roque de los Muchachos (ORM), at 2.2~km asl. The system
telescopes feature a mirror area of 8.5~m$^2$  and a focal
length of 5~m and are equipped with 271-pixel 
photomultiplier cameras. Based on
the multiple views obtained for each shower, the orientation
of the shower axis in space as well as the location of the
shower core can be determined. Compared to single IACTs,
stereoscopic IACT systems provide superior angular resolution,
energy resolution, and background rejection
\cite{performance,trigger_paper,mrk501}.
\begin{figure}[tb]
\begin{center}
\mbox{
\epsfxsize12.0cm
\epsffile{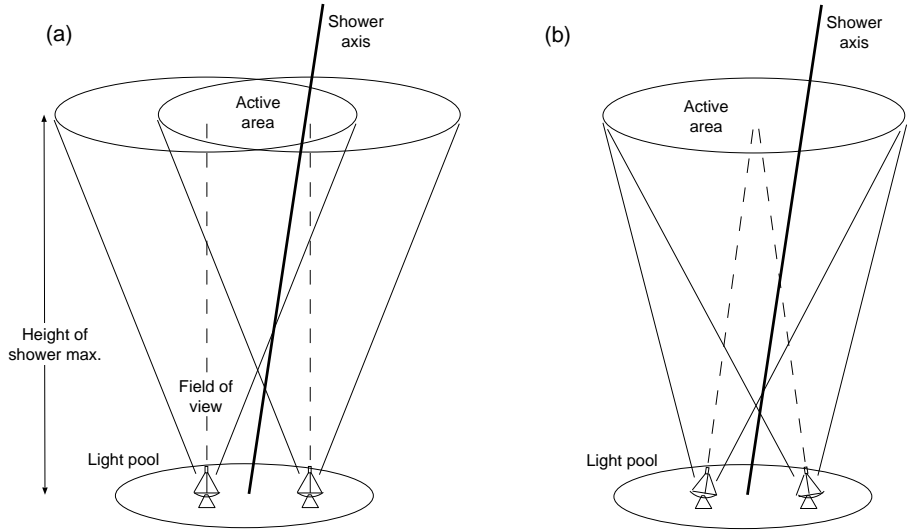}}
\end{center}
\caption
{Illustration of the geometry with parallel (a) and 
canted (b) telescopes. The dashed lines show
the optical axes of the telescopes. The ``active area'' indicates 
where the shower maximum can be observed in both telescopes.}
\label{scheme}
\end{figure}
During typical observations with the HEGRA IACT system, the
optical axes of all telescopes are parallel and either point
directly to the source, or -- in the so-called wobble mode --
at a point displaced by $\pm 0.5^\circ$ in declination 
relative to the source.
In the latter case, the rate in a region displaced by the
same distance from the optical axis, but in the opposite 
direction, is used to estimate off-source background rates.

With all telescopes pointing in exactly the same
direction, both the operation
of the system and the data analysis are simplified,
but one may wonder if the detection characteristics could not
be improved by canting the telescopes towards each other,
such that their optical axes intersect roughly at the 
height of the shower maximum. Such an alignment of 
telescopes would guarantee that the most luminous 
region of an air shower is optimally viewed by all
telescopes simultaneously.

The two alternatives are shown in Fig.~\ref{scheme},
which also serves to illustrate the trigger characteristics
of IACT arrays.
To first approximation, an individual IACT will
 trigger on an air 
shower if two conditions are fulfilled \cite{koehler}: 
\begin{enumerate}
\item
the telescope has 
to be located within the light pool of the shower, with its
typical radius of about 120~m, and 
\item
the shower maximum has to
be within the field of view of the camera. 
\end{enumerate}
For the HEGRA
telescopes, with their $4.3^\circ$ field of view,
the latter condition implies that the shower maximum --
at TeV energies typically located 6~km above the
telescopes -- should be within 225~m from the optical 
axis of the telescope. For showers propagating
parallel to the optical axis, the second condition is
automatically fulfilled, once a telescope is within
the light pool of the air shower. 
The camera field of view adds an additional constraint only
for showers at angles of more than $1^\circ$ relative to the 
optical axis (at least as far as triggering is 
concerned -- to avoid truncation of images, one
may want to require in addition in the subsequent
image analysis that the centroid 
of the image is at least $0.5^\circ$ away from the edge
of the field of view; this will still
result in a field of view of about 175~m radius at the shower maximum). 
Therefore, canting of telescopes is most likely not an
issue for studies of point sources near the center
of the field of view; it may however be important for
the observation of extended sources as well
as for surveys of larger areas of the sky, where it is
important to maximize sensitivity over a large solid angle.
For two telescopes, the situation is relatively obvious
from Fig.~\ref{scheme}: for parallel pointing, the
range of locations of the shower maximum, and hence the
accessible solid angle, is much more restricted than
for canted telescopes.
For three or more telescopes, the conclusion depends
on the locations of the telescopes and the trigger
conditions; parallel pointing will reduce the solid
angle for a coincidence of all $N$ telescopes, but may
increase the angle if only 2 out of $N$ telescopes are
required in the trigger and in the subsequent analysis.

\begin{figure}[htb]
\begin{center}
\mbox{
\epsfxsize11.0cm
\epsffile{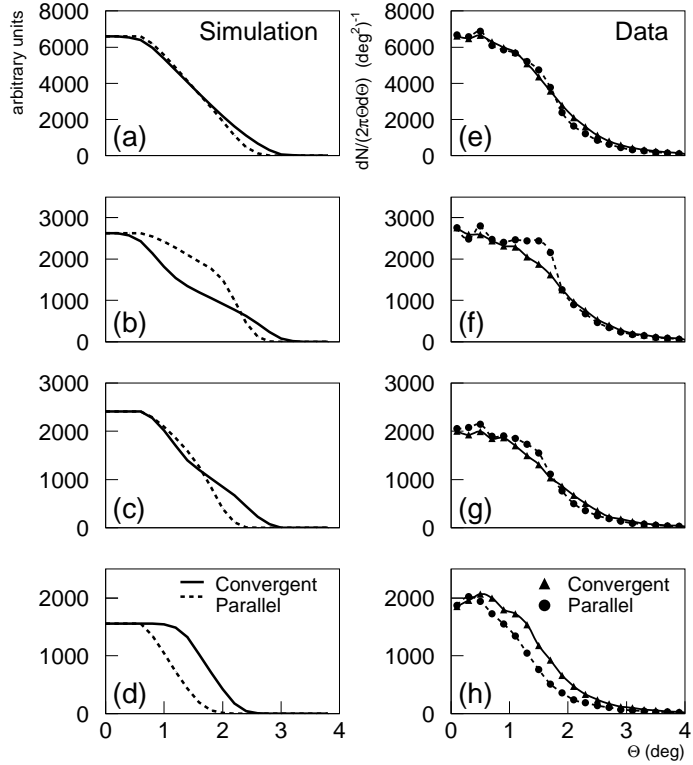}}
\end{center}
\caption
{Left: simple geometrical model for the number of
detected showers as a function of the angle between
the shower axis and the axis of the telescope system,
for parallel axes of the telescopes (dashed line) and
for canted telescopes (full line). (a) Total rate,
(b) 2-telescope events, (c) 3-telescope events,
(d) 4-telescope events. Right: experimental detection
rates for cosmic-ray showers, for all events (e), 2-telescope events (f),
3-telescope events (g) and 4-telescope events (h),
for parallel (points) and canted alignment of
axes (triangles).}
\label{rates}
\end{figure}

Estimates of detection rates were
carried out for the actual geometry of the 
HEGRA IACT system,
with telescopes located at three of the four corners of a square
of about 100~m side length, and another telescope located
at the center of the square (the remaining corner
telescope had at that time -- summer 1998 -- an older camera
and was not yet included
in the IACT system). The rate estimates were based on the 
simplified model discussed above, assuming a radius of the
light pool of  about 120~m and a usable field of view
(without edge distortion of the images) of $3.6^\circ$.
Two cases were compared: 1) parallel optical axes of
all telescopes, and 2) telescopes canted such that the 
axes intersect in a height of 6~km, with the nominal
pointing of the IACT system defined as the pointing of the
central telescope.
The results -- event rates for a given source flux and
observation time as a function of the distance of the source
from the optical axis of the central telescope -- are
shown in Fig.~\ref{rates}(a)-(d). The simulations indicate almost
identical total detection rates for the two pointing modes
(Fig.~\ref{rates}(a)). However, for sources more than
$0.5^\circ$ from the optical axis of the system, the 
``convergent observation mode'' provides a significantly larger fraction
of four-telescope events (Fig.~\ref{rates}(d)). Since both the angular 
resolution and the cosmic-ray rejection improve with
the number of triggered telescopes, the convergent mode
is clearly favorable.
%

%
\begin{figure}[tb]
\begin{center}
\mbox{
\epsfxsize14.0cm
\epsffile{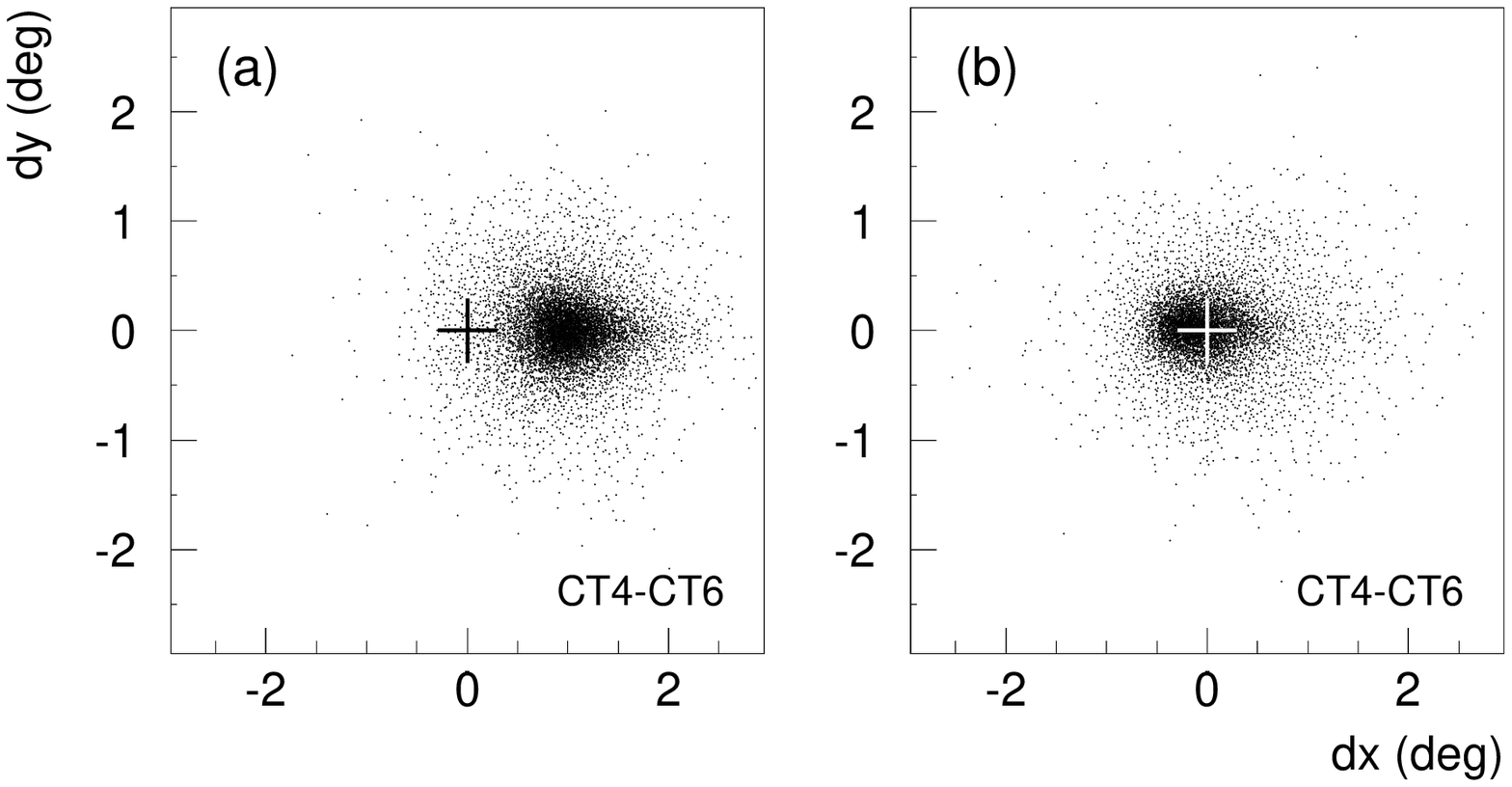}}
\end{center}
\caption
{Difference in the centroid positions of images in two cameras
 for parallel (a) and convergent
observations (b). The camera coordinate systems are rotated in a way 
that the x-axis points along the line connecting the telescopes.}
\label{dxdy}
\end{figure}
To verify these model predictions experimentally,
three hours of observations of the Crab nebula were
performed with canted telescopes. 
The cosmic-ray background
provides a uniform flux of particles and allows to
study detection rates as a function of the distance
to the optical axis of the IACT system. At last
qualitatively, these characteristics should be similar
for hadronic showers and for the more interesting
$\gamma$-ray induced showers. Limited observation
time and low counting rate prevented us from scanning
the field of view using the Crab nebula as a $\gamma$-ray
source. 

Fig.~\ref{dxdy} illustrates the effect of the canting:
the positions of the
images in the different cameras coincide, 
whereas for the parallel pointing mode
they are displaced by about 
$\delta \approx d/h \approx 1^\circ$ along the 
direction connecting the locations of the telescopes.
Here, $d$ is the spacing of the telescopes and $h$
the height of the shower maximum.
As a consequence, in convergent mode it is unlikely
that images in some of the telescopes are truncated;
either all telescopes have well-contained images,
or all images suffer from edge problems.

Due to slight differences in the weather conditions,
the telescopes trigger rates varied somewhat
between the observations taken in convergent
tracking mode, and the reference data set. Since
the pointing mode may influence the trigger rates,
one cannot simply normalize the data sets on the
basis of the raw trigger rates.
To derive the correct normalization factor, the field of view in
convergent mode was artificially truncated to 
$2.4^\circ$ diameter, and in the reference data set 
-- with parallel pointing --
the convergent pointing was mocked up by selecting
in software a $2.4^\circ$ field of view shifted 
by the canting angle. After these software trigger cuts, 
the detection rates can be compared, 
resulting in a $24 \pm 2\%$ correction.

The corrected cosmic-ray detection rates as a function of the angle relative
to the system axis are shown in Fig.~\ref{rates}(e)
for the total rate, and separately for 2-telescope (f), 
3-telescope (g), and 4-telescope
events (h). Here, only images within the central $3.6^\circ$
of the field of view were accepted, to exclude truncated
images. The observed pattern matches that predicted by
the simple model: very similar total rates, but a clear
enhancement of the 4-telescope rate at larger angles
in case of the convergent pointing mode, at the expense
of 2 and 3-telescope events. For the 4-telescope events,
the diameter of the 
effective field of view is increased by about 
$0.8^\circ$.

During the data taking in convergent mode, the Crab
nebula was positioned $0.5^\circ$ off the system optical
axis. Under these conditions, one would not expect 
any difference in detection rates between the two modes,
and indeed the Crab signals in the two modes are well
consistent within the statistical errors of about 20\%.


In summary, for stereoscopic IACT systems, the convergent tracking mode 
-- canting the telescopes towards each other such that
their optical axes intersect at the height of the shower 
maximum -- improves the detection capabilities in  particular
for sources near the edge of the field of view, and is
advised for observations of extended sources and for surveys.

{\bf Acknowledgments.}
The support of the German Ministry for Research 
and Technology BMBF and of the Spanish Research Council
CYCIT is gratefully acknowledged. We thank the Instituto
de Astrofisica de Canarias for the use of the site and
for providing excellent working conditions. We gratefully
acknowledge the technical support staff of Heidelberg,
Kiel, Munich, and Yerevan.

\end{document}